\documentclass[11pt]{article}
\usepackage[T1]{fontenc}
\usepackage[margin=1in]{geometry}
\usepackage{amsmath,amssymb,booktabs,graphicx,xurl}
\usepackage[hidelinks]{hyperref}
\usepackage[section]{placeins}
\newcommand{\SampleN}{240}
\newcommand{\JointN}{166}
\newcommand{\ObsN}{212}
\newcommand{\TrueEdges}{713}
\newcommand{\UnknownN}{74}
\newcommand{\QueueN}{51}
\newcommand{\OutsideN}{23}
\newcommand{\Coverage}{69.17\%}

\title{Architectural Debt Enforcement under Incomplete Observation:\\A Controlled Study of Ratchets and Evidence Qualification}
\author{Zhiheng Zhang\\University of Melbourne\\\small ORCID: 0009-0005-9005-9931}
\date{30 October 2026}
\begin{document}
\maketitle
\begin{abstract}
Architectural debt gates must distinguish an allowed change from an incompletely
observed one. We study a decision contract separating observation qualification,
prohibited relation identity, inherited debt and authorized baseline updates.
Across \SampleN{} fixed files from four repositories, the current observer and
programmatic reference jointly qualify \JointN{} files (\Coverage{}), all with
exact reference-relative edge agreement; \UnknownN{} remain reference-UNKNOWN.
Coverage varies from 13/60 to 59/60 by repository. On 480 constructed state
scenarios sharing 60 source anchors, both the candidate ratchet and mature
dependency-cruiser baseline match every expected outcome, whereas static and
count-only baselines match 420 and 360. These results do not establish candidate
superiority. Preserved failures include nine additional false-PASS probes and
120 unqualified full-tree timing attempts. A subsequent generic binding layer
rejects 54 specified mismatches, accepts 54 current controls and permits legal
repetition. Two bounded package scopes support 40 scenarios and 48 fresh-process
functional checks; their costs remain tied to the measured implementations.
Diagnostics explain residual uncertainty through type/runtime relations,
installation context and generated-artifact provenance. TypeScript 5.9.3 and
6.0.2 yield identical outputs on ten paired emission witnesses, but neither that
agreement nor verified release provenance changes the frozen UNKNOWN labels.
The contribution is a controlled account of evidence-qualified enforcement and
its limits, not language completeness, natural maintenance effectiveness or a
new shrinking-baseline algorithm.
\end{abstract}
\section{Introduction}
A dependency gate can prevent a prohibited architectural relation only if its
observation represents the relevant source. An empty violation list can mean
that no forbidden relation exists, but also that a scan omitted a file, a module
expression exceeded the observer's model, or evidence came from a stale source
state. Existing debt creates another distinction: preserving its count does not
preserve its identity. Removing one violation while introducing another can keep
the total unchanged, and a static permission for an inherited relation can
allow that relation to return after repair.

Shrinking known-violation baselines already address the latter state problem in
mature tools. The unresolved premise is whether a particular decision is backed
by a qualified, current observation under an authorized policy and debt state.
This premise is easy to hide when evaluation reports agreement only on files
that its reference can resolve. In our fixed census, perfect agreement within
the jointly qualified subset coexists with \UnknownN{} reference-UNKNOWN files.
The main empirical question is therefore what those decisions establish and
what remains outside their evidence boundary.

We report a controlled mechanism study with three contributions. First, we
make the decision contract explicit: observation qualification, relation
identity, debt retirement and state authority have distinct roles. Second, we
retain and analyze observation and evidence-binding failures, and evaluate a
generic current-context layer that can wrap a mature dependency tool. Third,
we trace residual uncertainty within fixed source snapshots to specific
type/runtime, installation and artifact-context questions. This diagnosis is
empirical and bounded; it is not a universal taxonomy of resolver failures.

The paper organizes the evidence around four reporting questions:
\begin{enumerate}
\item RQ1: What coverage and reference-relative agreement are established,
      and how do they vary across the sampled repositories?
\item RQ2: Which constructed changes distinguish count, static-identity and
      shrinking-debt semantics, including mature-tool alternatives?
\item RQ3: When do missing or mismatched inputs fail qualification, and what
      does generic current-context binding establish?
\item RQ4: What successful operation and cost are demonstrated on explicitly
      bounded, functionally checked scopes?
\end{enumerate}
Coding agents motivate the need for protected governance, but no natural agent
trajectory or productivity outcome is measured. Controlled scenarios, repeated
source anchors and post-exposure diagnostics are identified as such. A negative
result remains part of the study even when a later engineering repair succeeds.

\section{Related Work}
Software reflexion models compare a mapped source model with a high-level
structural model and make their agreement and disagreement visible\cite{reflexion}.
Architecture reconstruction also depends on the goals, inputs and abstraction
process used to recover a view\cite{reconstruction}. These precedents matter
here because a dependency graph is an observation under a model, not an
unqualified account of all runtime behavior. Our study concentrates on whether
such an observation is admissible for an enforcement decision and a debt update.

Existing tools already support inherited violations. ArchUnit's
\texttt{FreezingArchRule} stores existing violations and reduces the stored set
when violations are repaired\cite{archunit}. Dependency-cruiser 18.7.0 exposes
known-violation baselines and a \texttt{shrink-only} update mode that removes
existing entries without adding new ones\cite{cruiser}. We consequently treat
shrinking identity as an established comparator capability. Our state experiment
does not establish a novel set algorithm or a unique gate advantage. The
Boundaries dependency rule evaluates configurable module policies\cite{boundaries};
its applicable outcomes and capability gaps are reported separately instead of
scoring a missing duty as an error.

TypeScript explicitly distinguishes import elision from verbatim module
syntax\cite{typescript}. This is directly relevant to architectural observation:
the presence of a source import and the presence of an emitted import need not
coincide. Package provenance addresses a different question. npm's provenance
documentation describes links between an artifact, its source and its build
environment\cite{npm}; such a link does not by itself reproduce the build or
select the admissible graph universe of this study.

Longer-horizon agent evaluation, exemplified by SWE-CI's iterative maintenance
tasks and CI-based protocol\cite{sweci}, asks about code evolution through actual
agent work. Our constructed changes instead isolate enforcement mechanisms.
They can reveal a counterexample or exercise a state contract but cannot estimate
an agent's maintenance effectiveness. This positioning is a focused comparison
with relevant primary sources, not a systematic literature review.

\section{Decision Model and Trust Boundary}

Let $E_t$ be the observed internal relation set, $P$ a fixed set of prohibited
identities, and $B_{t-1}$ the authorized inherited debt. Identities use normalized
directed source--target paths under the fixed relation semantics; import form
is retained as evidence and is part of identity only when the locked policy
specifies it. Unsupported or unresolved observations prevent qualification
under the recorded model rather than implying an allowed relation.
With qualification $Q_t$,
the decision and accepted shrinking update are
\begin{align}
 N_t &= (E_t\cap P)\setminus B_{t-1},\\
 V_t &=\begin{cases}\textsc{QFail}&\neg Q_t,\\
 \textsc{Reject}&N_t\ne\varnothing,\\\textsc{Accept}&\text{otherwise},\end{cases}\\
 B_t&=\begin{cases}B_{t-1}\cap(E_t\cap P)&V_t=\textsc{Accept},\\B_{t-1}&\text{otherwise}.\end{cases}
\end{align}
A count gate can miss compensation: one old relation disappears while a new
prohibited identity replaces it. A static identity baseline can permit
reintroduction after deletion. Given a qualified graph that actually contains
the relation, the shrinking rule prevents both constructions. This follows
from set membership: deletion removes the identity from the accepted baseline,
so later reintroduction belongs to $N_t$. The premise is essential; this
argument does not prove that an observer sees every relation.

\subsection{Current context, cached observation and state commit}
The supplementary trusted runner independently derives a context digest from
actual source and overlay bytes, scan scope and relation semantics, current policy,
authorized baseline contents and epoch, observer/resolver identity, and
configuration. Evidence must be internally consistent \emph{and} match that
current context and the runner's own observation. Rehashing an arbitrary
payload alone is insufficient. The trusted graph cache is keyed by the
source/scope/observer/configuration context; policy and baseline are rechecked
for each decision. Identical-state repetitions and legal cached observations
remain usable, with fresh gate receipts rather than a nonce that disables
idempotence.

AST observation reads an immutable in-memory materialization. Before an
accepted debt update, the runner rechecks inputs and the baseline epoch, then
uses compare-and-swap on a private in-process state store. Policy rejection
and qualification failure leave debt unchanged. The native dependency-cruiser
adapter rereads its own files and computes its own policy outcome; it does not
consume an A 2 or oracle graph as its native input. Its generic wrapper checks
current context around that execution. This bounds ordinary injected races,
not a hostile operating system or durable distributed transaction.

The trust assumption is a protected runner, policy and baseline authority.
An agent may change ordinary source but cannot legitimately create the expected
context or authorize its own debt. The prototype does not implement a signature
infrastructure, prove parser correctness, protect a malicious host, or certify
an external dependency supply chain. Package runtime locks and archive integrity
are recorded separately. Input binding is orthogonal to observation coverage:
a correctly bound incomplete graph can still be semantically inadequate.

\section{Study Design}

\subsection{Immutable original study and post-study repair}
The author-approved original protocol fixes four pinned external repositories
and sample quotas (Appendix~\ref{app:repro}). E1 uses 240 files, 40 random plus 20 distinct
stress files per repository. E2 applies eight mutation families to 20 source
anchors in each of Boss, Vue and tRPC:480 scenarios but only 60 repository/source
anchors. The linked deletion/reintroduction steps form one scenario. E3 has
six fault families of 10 constructions and 60 valid controls. E4 has 30 timed
attempts per applicable configuration, with warmups separate. All failed and
UNKNOWN records remain in their original files.

Diagnosis exposed 9 false PASS outcomes outside the narrow primary E3 apparatus.
The first isolated repair changed lexical observation to a qualified AST path
and improved explicit resolution. Original, failed and corrected attempts are
retained. The repair's 41-case set included a newly exposed grammar failure that
was subsequently repaired; its final success is regression evidence. The
separate 16-case extension has its own lock and exposure record. We do not fold
these probes into the primary E3 denominator.

\subsection{Prospective supplementary qualification and reference checks}
The supplement was specified after the preceding failures were known. A fresh
non-authoring evaluation context generated 120 qualification challenges:
10 families, 6 matched pairs each. The observer code was frozen before its first
run. The authoring context did not inspect the case contents before recording
that run. Language-invalid source and valid but unsupported model inputs are
separated. Once outcomes were used for repair, those cases became exposed
regressions. A separately locked 24-case extension is reported with its own
first-run timing; none of these designed cases estimates natural fault
incidence or represents human annotation.

We re-evaluate the original 240 E1 files without altering their old labels.
A separately versioned reference uses Babel traversal and the official
TypeScript module resolver over an exact, case-sensitive tracked virtual
filesystem and explicit root configuration. Node resolution witnesses provide
bounded checks on disagreements. Unsupported bundler conditions, declarations
and missing generated targets remain UNKNOWN. Independent review identified
an initially incomplete reference binding check; its outputs remain, and a
corrected reference version and difference ledger are supplied. Shared library
and configuration assumptions remain a threat even with separate traversals.

\subsection{Binding faults and valid controls}
We prescribe nine binding fault classes with six constructions each: stale
source, changed scope, changed policy, stale baseline epoch, replaced debt with
a recomputed self-hash, replaced observation, partial evidence marked complete,
post-scan source change, and replay after debt retirement. Each has a current,
diversified legal control. Six additional repeat/cache pairs test idempotence.
Only the previously demonstrated stale-evidence gap is treated as an established
old-system failure; other constructions are prospective supplementary tests,
not retrospectively claimed exploits. Controls are still templates, not a
representative sample of production changes.

\subsection{Fixed package scopes and two comparator layers}
Before new gate outcomes we selected Vue's complete shared-package source
scope and Fastify's published runtime entry/lib scope. We retain all original
all-tracked-source failures. The first pilot failed: build boundary inputs and
runtime acquisition were incomplete. The corrective build materialization
follows the deterministic tracked source closure without shrinking or replacing
the scan scopes. Cross-scope edges remain visible, while unscanned outgoing
relations are not claimed observed.

For each project, five state-change constructions and four deterministic seeds
produce 20 scenarios. They exercise a normal change, an allowed relation, debt
deletion, linked reintroduction, and identity compensation. These 40 units use
only two original source anchors; they are finite mechanism demonstrations,
not 40 independent maintenance tasks. Policies and lab helper modules are
explicitly experimental, not endorsed maintainer architecture.

A final audit found that repeated same-process smoke calls could reuse Node
module caches. We retained those results and re-executed all 48 locked source
states in separate processes and fresh directories; these latter records govern
functional validity. Functional evidence is actual Fastify HTTP injection and TypeScript-to-ESM
Vue shared API assertions in the recorded build context. It is stronger than
syntax-only controls but is not the full maintainer build/test suite. The
native runtime dependencies were acquired in a fresh directory and verified
against the locked package archive integrity; initial acquisition failures are
retained.

Layer A supplies the same independently reference-qualified graph to state
evaluators, including the actual native baseline diff/reporter with an explicit
research graph adapter. Layer B uses each tool's own observation: original
lexical shrinking, the AST/bound candidate, native dependency-cruiser shrink-only,
and dependency-cruiser plus the same generic binding layer. Native observation,
qualification and final policy outcome are separate. Native tools receive N/A
for evidence duties they do not claim, not an artificial failure.

\subsection{Analysis, cost and resource accounting}
We report complete censuses, coverage, UNKNOWN reasons, edge agreement and file
exactness with explicit reference versions. Original E2 paired contrasts are
resampled by source anchor and by mutation family. Case-level tiny p-values are
not a principal result. Degenerate anchor intervals can follow from applying
the same eight constructions to every anchor; they do not establish population
precision. Family resampling exposes a different and small design universe.

New cost measurements use frozen qualified package snapshots, 30 fresh-process
repetitions per system/project and separate warmups, with deterministic order
rotation. Method wall/CPU and whole-process footprint have distinct meanings.
The common harness contributes to peak RSS; job committed memory is not RSS.
OS caches remain warm, and the host is not a dedicated benchmark machine.
Different duties make these a cost decomposition, not universal speed superiority.
We retain original rejection-path costs under the original observer only.

The fixed experimental allowance is cumulative: eight CPU-affinity slots,
16 GiB job committed memory, at most 120 GiB owned storage, and 12 hours of
experimental computation. Windows admission and periodic storage checks
replace a hard disk quota. The present integration imports saved records and
does not rerun experiments, change samples or reset prior charges.

The reporting questions below reorganize the approved observation, state and
qualification/cost questions; they do not amend the original three-RQ protocol.
RQ4 groups the already specified scoped supplement and cost evidence. Post-study
repairs and diagnostics remain separate from original E1--E4 measurements.
\begin{table}[htbp]
\centering\small
\begin{tabular}{lrrr}
\toprule
Record phase & Observer & Reference & Gate / duty \\
\midrule
Original & O0 & Original & Original \\
V2 supplement & O1/O2 & R2/R3 & G2 \\
V3 reconciliation & O3 & R4/R5 & G2 replay \\
V4--V7 diagnostics & Unchanged O3 & Unchanged R5 & No new gate run \\
V8 integration & No execution & Saved records & Editorial checks \\
\bottomrule
\end{tabular}
\caption{Paper phases do not rename measured implementations. V2 cost and functional evidence remain attached to their original locks.}\label{tab:versions}
\end{table}

\section{Results}
\subsection{RQ1: Coverage precedes conditional agreement}
The current O3 observer qualifies \ObsN{} of \SampleN{} files; the R5 reference
qualifies \JointN{}. Their intersection contains \JointN{} files, all edge-exact
relative to that reference, with \TrueEdges{} true-positive edges and zero
false positives or false negatives. Table~\ref{tab:coverage} puts the whole
denominator first. The \Coverage{} coverage is not an accuracy estimate over
all files, and the observer's more permissive qualification does not certify
the additional files as correct. Fastify contributes 47 of the \UnknownN{}
reference-UNKNOWN files, while tRPC contributes one. This heterogeneity limits
interpretation of a single pooled conditional score.
\begin{table}[htbp]
\centering\small
\begin{tabular}{lrrrr}
\toprule
Repository & Files & Joint exact & Ref. UNKNOWN & Coverage \\
\midrule
fastify/fastify & 60 & 13 & 47 & 21.67\% \\
trpc/trpc & 60 & 59 & 1 & 98.33\% \\
vitejs/vite & 60 & 42 & 18 & 70.00\% \\
vuejs/core & 60 & 52 & 8 & 86.67\% \\
Total & 240 & 166 & 74 & 69.17\% \\
\bottomrule
\end{tabular}
\caption{Current O3/R5 reference-relative coverage. Each repository retains 40 random and 20 stress files. UNKNOWN is not an error label.}\label{tab:coverage}
\end{table}

The remaining \QueueN{} reference/observer disagreements are a diagnostic queue
within the \UnknownN{} reference-UNKNOWN files; \OutsideN{} other UNKNOWN files
lie outside that queue. Reconciliation closed ten of the original 61 queued
disagreements: five became jointly qualified, while five merely obtained equal
graphs with both systems still UNKNOWN. The engineering changes correct a
trailing-slash directory-host defect, normalize a parent-relative package main
before checking containment, and recognize explicit tracked asset targets
without guessing extensions. Prior labels and outputs are retained separately.
Subsequent diagnostics explain UNKNOWN cases but do not alter this census.

\subsection{RQ2: Identity and state, without a mature-tool advantage}
Table~\ref{tab:state} reports all original E2 methods. Candidate shrinking and
dependency-cruiser both match 480/480 constructed outcomes. Static identity
matches 420/480 and count-only debt 360/480. A compensating change defeats the
count rule; deletion followed by reintroduction defeats the static baseline.
These constructions demonstrate the stated set semantics when the relevant
graph is observed. They do not show that a particular implementation observes
every dependency. The qualification ablation also matches 480/480, so this
corpus alone provides no empirical advantage for qualification.
\begin{table}[htbp]
\centering\small
\begin{tabular}{lrrr}
\toprule
Method & Correct & Applicable & N/A \\
\midrule
Candidate shrink & 480 & 480 & 0 \\
Count only & 360 & 480 & 0 \\
Qualification ablation & 480 & 480 & 0 \\
Sensor ablation & 420 & 480 & 0 \\
Static identity & 420 & 480 & 0 \\
Dependency-cruiser & 480 & 480 & 0 \\
Boundaries & 420 & 420 & 60 \\
Boss current & 140 & 160 & 320 \\
\bottomrule
\end{tabular}
\caption{Original E2: 480 constructed scenarios, 60 source anchors. N/A is retained; the qualification ablation ties the candidate.}\label{tab:state}
\end{table}

The 480 units reuse 60 source anchors across eight mutation families. In each
project, the paired difference from static and count baselines is 0.125 and
0.250. Saved family-resampling descriptive intervals are [0, 0.375] and
[0, 0.500]; anchor-resampling intervals degenerate because each anchor repeats
the same families. The contrast with dependency-cruiser is zero under both
resampling units. These results describe a constructed design, not equivalence
or population-level precision. N/A outcomes for Boundaries and the original
Boss evaluator remain outside their applicable denominators.

\subsection{RQ3: Qualification failures and current-context binding}
Original E3 rejects 60/60 injected faults and accepts 60/60 valid controls, but
the controls all use \texttt{export const allowed=1;}. Nine additional probes
produce nine false PASS outcomes. The isolated repair subsequently passes its
41-case regression set, which includes an exposed-and-repaired grammar failure,
and a separately locked 16-case extension. These are distinct evidence sets;
none replaces the primary E3 denominator.

In the supplementary 120-case qualification challenge, immutable O1 first
matches 114 cases and falsely qualifies six owned-package targets. The first
conservative repair trades these errors for six false refusals. Explicit
tracked-main resolution then yields 120/120 on the now-exposed regression set.
A separately locked extension matches 24/24 on its first run. Later repetitions
are regression checks, not additional unseen validation.
\begin{table}[htbp]
\centering\small
\begin{tabular}{lrrr}
\toprule
Stage & Matches & False qualification & False refusal \\
\midrule
O1, first run & 114/120 & 6 & 0 \\
Intermediate, exposed & 114/120 & 0 & 6 \\
O2, regression & 120/120 & 0 & 0 \\
Locked extension, first run & 24/24 & 0 & 0 \\
\bottomrule
\end{tabular}
\caption{V2 model-qualification challenges. Repairs use exposed cases; subsequent repetitions of the extension are not unseen validation.}\label{tab:qualification}
\end{table}

The generic binding layer yields the specified outcomes for all 54 mismatches
and 54 current controls (Table~\ref{tab:binding}). Controls have distinct source
hashes but remain templated constructions. Six legal repeat/cache groups succeed
with fresh receipts and stable authorized state. Mismatches include source,
scope, policy, baseline epoch, debt replacement, observation replacement,
incomplete scans, injected post-scan changes and replay after debt retirement.
Each failed decision preserves the authorized debt state. This supports the
tested protected-runner contract, not hostile-host security.
\begin{table}[htbp]
\centering\small
\begin{tabular}{lrr}
\toprule
Construction & Expected outcomes & Total \\
\midrule
Mismatched evidence: QFAIL & 54 & 54 \\
Current controls: ACCEPT & 54 & 54 \\
Legal repeat/cache groups & 6 & 6 \\
\bottomrule
\end{tabular}
\caption{V2 generic binding under the protected-runner assumption. Failed decisions preserve authorized debt.}\label{tab:binding}
\end{table}

\subsection{RQ4: Bounded operation and qualified-snapshot cost}
The complete Vue shared-package and Fastify runtime scopes contain respectively
16 and 33 source files, plus three deterministic helper files each. After a
preserved failed pilot, the 40 locked scenarios across two source anchors pass
their limited functional assertions in 48 fresh-process source states. The
assertions exercise Fastify HTTP injection and Vue shared APIs after the
recorded TypeScript-to-ESM build. They are not full maintainer CI suites.

In the shared-graph layer, shrink and the native reporter adapter match 40/40,
static identity 32/40 and counts 24/40. The native reporter adapter initially
matched only 24/40 because of inconsistent baseline-metadata projection; that
failed research-adapter attempt is preserved. Under each tool's own observation,
all four methods match the 40 terminal decisions: 24 ACCEPT and 16 REJECT.
The ordinary native tool has N/A for trusted evidence-binding duties; the
wrapped native method performs those additional checks.
\begin{table}[htbp]
\centering\small
\begin{tabular}{lrr}
\toprule
Layer / method & Matches & Scenario total \\
\midrule
Shared graph: shrink & 40 & 40 \\
Shared graph: native reporter & 40 & 40 \\
Shared graph: static & 32 & 40 \\
Shared graph: count & 24 & 40 \\
Own graph: O0\_LEXICAL\_SHRINK & 40 & 40 \\
Own graph: AST\_BOUND\_V2 & 40 & 40 \\
Own graph: DC\_NATIVE & 40 & 40 \\
Own graph: DC\_BOUND\_V2 & 40 & 40 \\
\bottomrule
\end{tabular}
\caption{V2 fixed Vue/Fastify scopes: two anchors, 40 scenarios and 48 fresh-process functional states. Native binding duties are N/A unless wrapped.}\label{tab:scope}
\end{table}

These successful scopes do not repair the original full-tree result. All 120
original candidate timing attempts complete as processes but fail qualification.
The original E4 additionally retains 60 Vite technical failures across native
comparator configurations. A successful exit or recorded latency cannot be
counted as qualified architectural governance. Later AST repair still leaves
the four original full-tree configurations unqualified.

Only the qualified V2 package snapshots support the successful-operation cost
table. All 240 timed runs ACCEPT, with eight warmups separate. The measured
environment is Windows 11 Home build 26200 on an Intel Core i7-14650HX,
restricted to eight processor-affinity slots and 16 GiB job committed memory,
with Node 24.14.0, TypeScript 5.9.3, Babel parser 7.28.5 and dependency-cruiser
18.7.0. Fresh processes retain warm OS caches on a non-dedicated host.
\begin{table}[htbp]
\centering\small
\begin{tabular}{lrrr}
\toprule
Project / configuration & N & Median (s) & P95 (s) \\
\midrule
fastify / AST\_BOUND & 30 & 0.1258 & 0.1591 \\
fastify / DC\_BOUND & 30 & 0.4174 & 0.4636 \\
fastify / DC\_NATIVE & 30 & 0.2602 & 0.3093 \\
fastify / O0\_LEXICAL & 30 & 0.0464 & 0.0551 \\
vue / AST\_BOUND & 30 & 0.1495 & 0.1896 \\
vue / DC\_BOUND & 30 & 0.3336 & 0.3668 \\
vue / DC\_NATIVE & 30 & 0.1733 & 0.1921 \\
vue / O0\_LEXICAL & 30 & 0.0097 & 0.0575 \\
\bottomrule
\end{tabular}
\caption{V2 frozen implementations, TypeScript 5.9.3, fresh processes and warm OS caches on Windows. All 240 timed outcomes ACCEPT; eight warmups are separate. Duties differ; these are not O3/R5 or TypeScript 6.0.2 timings.}\label{tab:cost}
\end{table}

The generic native wrapper adds checking and native reevaluation, so the rows
decompose differing duties rather than demonstrate universal speed superiority.
CPU, footprint and saved intervals remain in the evidence records. V3 replays
40 terminal scenarios and 48 graph/state steps; it does not rerun the native
tools, functional tests or timings. Neither O3/R5 nor TypeScript 6.0.2 inherits
the older cost labels.

\section{Explaining Residual Uncertainty}
The residual diagnosis explains RQ1's coverage and RQ3's qualification boundary.
It does not introduce a new benchmark or redefine an admissible relation.
The \QueueN{} queued cases comprise 45 declaration/runtime-model cases and six
workspace-context cases. The complete UNKNOWN census also retains the
\OutsideN{} cases outside that queue. Source hashes, witness records and unresolved
human questions remain linked by case identity.

\subsection{Type targets, runtime paths and import elision}
Across the 45 declaration cases, 59 distinct case/specifier/form tuples reproduce
the recorded TypeScript declaration targets. Forty JS/MJS files contribute 41
mode-appropriate Node path probes, all pointing to existing tracked runtime
files. These are path-resolution witnesses, not module execution or proof of
export equivalence. The remaining five files are TypeScript type tests with
18 tuples: five counterfactual ESM paths name existing files, whereas 13 have
no corresponding runtime file. Returning an ESM URL alone is insufficient to
claim that a module can load.

Single-file emission supplies a more direct test of the missing-path explanation.
The same five source hashes are passed to TypeScript 5.9.3 and 6.0.2 through
\texttt{transpileModule}, with target ES2022 and module ESNext. Only
\texttt{verbatimModuleSyntax} varies between false and true. Babel inventories
import declarations in emitted output as an independent syntax parser, not an
independent semantic oracle. With false, five of 18 tuples survive and all 13
missing-path imports disappear. With true, all 18 survive. All ten paired
cross-version outputs are byte-identical (Table~\ref{tab:typescript}); these are
five sources times two modes, not ten independent source cases.
\begin{table}[htbp]
\centering\small
\begin{tabular}{lrrrr}
\toprule
Compiler & Verbatim & Files & Imports retained & Diagnostics \\
\midrule
5.9.3 & false & 5 & 5/18 & 0 \\
5.9.3 & true & 5 & 18/18 & 0 \\
6.0.2 & false & 5 & 5/18 & 0 \\
6.0.2 & true & 5 & 18/18 & 0 \\
\bottomrule
\end{tabular}
\caption{Saved single-file emission witnesses, ES2022/ESNext. The same five sources yield ten cross-version output pairs, all byte-identical. Diagnostics are not full type checks; emitted JavaScript was not executed.}\label{tab:typescript}
\end{table}

This result narrows the explanation without selecting a relation layer. The
maintainer type-test configuration declares \texttt{noEmit}, extends
\texttt{fastify-tsconfig} under a \texttt{\^{}3.0.0} dependency and uses TSTyche.
Its effective external configuration and exact historical installation remain
unresolved. TypeScript 6.0.2 is the lower bound of the declared
\texttt{\~{}6.0.2} range, not proof of the installed version. Our deliberate
emission is not the maintainer's type-test workflow. Zero transpile diagnostics
does not establish full-program type correctness, and generated JavaScript was
not executed. A missing counterfactual path therefore need not become an emitted
import, but no original UNKNOWN is made qualified by this observation.

\subsection{Local installation and template contexts}
Three of six workspace cases have pinned manifest and lockfile evidence for
specific local packages. In nine controlled arms, each unchanged importer is
tested with the package absent, an exact copied target present, and changed
content at the same target path. All outcomes match the predeclared expectation.
The last arm establishes dependence on loaded content; it does not test selection
between alternate paths. These minimal copied layouts do not reproduce pnpm
link/file semantics, symlinks, peer dependencies, a historical installation or
the full Vite loader.

Two other cases declare generated distribution entries absent from the tracked
snapshot. Replacing them with source entries would change the evaluated relation.
The sixth case is a scaffold. Pinned create-vite source copies a template into
a destination, changes its manifest and installs there or prints installation
instructions; a beta branch can also change the Vite version range. This
supports a generated-project installation context as an intended use. It does
not show which choices or installation occurred for the sampled file, and no
generator was executed. The six files contain seven specifiers, not seven
independent installation experiments.

\subsection{Release artifacts and source provenance}
Exact-version npm archives contain the advertised Vite and legacy-plugin
distribution entries missing from Git. Archive integrity establishes agreement
with saved registry digests; it is separate from source identity. Both version
metadata records omit \texttt{gitHead}, which initially leaves source identity
unestablished rather than demonstrating a mismatching commit.

Later saved verification supplies an authenticated provenance link. Vite 7.3.7
names the frozen commit; the legacy plugin 7.2.1 names a different commit
(Table~\ref{tab:provenance}). The latter is a valid release-source association
to another source state, not a signature-verification failure. Four verified
attestations comprise publish and SLSA statements for two packages, not four
independent package cases. Offline verification accepts two original bundles
and rejects four targeted negative controls; this small control set is not a
general security benchmark.
\begin{table}[htbp]
\centering\small
\begin{tabular}{lrrr}
\toprule
Distribution & Verified source & Frozen match & Build reproduced \\
\midrule
vite 7.3.7 & 284f6b58d6b5 & Yes & No \\
vitejs-plugin-legacy 7.2.1 & 48139b4828cb & No & No \\
\bottomrule
\end{tabular}
\caption{Verified release-source association is distinct from reproducing a build or admitting generated files to the frozen source graph.}\label{tab:provenance}
\end{table}

Neither association is a locally reproduced build. Selected legacy workflow,
source, manifest and configuration files differ from the frozen counterparts;
their hashes alone do not establish which differences affect emitted bytes.
Vite's narrow provenance subquestion is supported, while legacy retains a
different-source limitation. In both cases generated files remain outside the
frozen tracked-source graph. The unresolved decision is artifact admissibility
under that model, not merely whether a downloadable file exists.

\subsection{The complete UNKNOWN population}
The additional \OutsideN{} UNKNOWN files are not omitted from coverage. Five
have missing distribution targets and nine have package-import mapping evidence.
These features may overlap, so they neither partition the 23 nor imply a
remaining category of nine. Computed imports, declaration targets, indirect
\texttt{require} behavior and conditional context remain recorded. Across all
diagnostics, automatic closures are zero and the frozen labels are unchanged.
The evidence improves the questions for human review without manufacturing an
answer to a different architectural model.

\section{Discussion and Threats to Validity}
\subsection{What the decision contract contributes}
Debt retirement is straightforward set bookkeeping once a qualified observation
and authorized state are available. The empirical difficulty is establishing
those premises. A bound but incomplete graph can still miss a forbidden relation;
a complete graph from a stale source state can still justify the wrong decision.
Our preserved failures demonstrate that a narrow integrity test and a successful
process are insufficient substitutes. The generic wrapper separates these
obligations from the underlying policy engine, allowing a mature tool to retain
its own observation and state semantics.

This is a compositional engineering lesson, not evidence that the candidate
outperforms mature alternatives. Dependency-cruiser ties the candidate on the
original state scenarios; all own-observation methods tie on the limited scopes.
The qualification ablation's tie also limits the original corpus's diagnostic
power. The binding constructions add evidence about a separately specified
duty, but their result cannot establish universal fail-closed behavior.

\subsection{Observation validity and reference dependence}
The reference is programmatic and corrected after outcomes were observed. Its
separate traversal does not remove shared TypeScript-resolution assumptions or
substitute for human semantic adjudication. Joint exact agreement therefore
means agreement with this reference on its qualified subset. The large
repository variation and \UnknownN{} reference-UNKNOWN files preclude a general
language-wide precision, recall or accuracy claim. The fixed random/stress
allocation is descriptive of this design, not a probability sample of software.

The diagnosis separates engineering defects from absent semantic or contextual
evidence. The directory-host and path-normalization repairs have narrow
independent witnesses; declaration/runtime choices and missing build contexts
are not automatically Windows defects. Successful path probes, copied-package
arms, emission comparisons and provenance verification answer different
questions. None authorizes changing a frozen relation definition. The resulting
categories describe these snapshots and may not exhaust other ecosystems.

\subsection{Construct validity, exposure and generalization}
Mutation families deliberately target count compensation and reintroduction.
Reused anchors, seeds and linked steps do not produce independent maintenance
tasks. The limited functional assertions cannot replace maintainer test suites,
and two successful package scopes do not erase four unqualified full trees.
Post-exposure repairs establish regressions; the first locked extension is
reported separately and is still a designed case set. Neither interval estimates
nor perfect matches support natural defect-incidence or agent-productivity claims.

Current-context binding assumes protected policy, runner and state authority.
The private in-process epoch/CAS store is not distributed durability. A malicious
operating system, compromised parser or undetected change-and-restore sequence
during native reads remains outside the guarantee. Cryptographic release
verification addresses artifact association, not these runtime trust assumptions.

Cost observations come from one non-dedicated Windows host with warm OS caches.
Configurations perform different duties, and whole-process footprint includes
the common harness. Job committed memory is not RSS. The measurements establish
bounded operation under recorded versions, not Linux parity, scale-out behavior
or the speed of later repairs. All unfavorable attempts remain available.

For practitioners, the supported recommendation is to report qualification,
scope, current input identity and state authority alongside policy outcomes;
retain UNKNOWN and technical failures; and keep legal repeated observations
usable while checking the current policy and baseline. Human adjudication,
broader build-system integration and longitudinal agent studies are future work.

\section{Conclusion}
Architectural enforcement depends on more than a violation count or a successful
tool invocation. This study separates observation qualification, prohibited
identity, debt retirement and authorized state updates. Mature shrinking tools
match the candidate on the controlled state experiment, while preserved false
PASS and full-tree qualification failures limit any stronger governance claim.
A generic binding layer supports the specified protected-runner checks and
finite successful operation on two package scopes. The current census certifies
only reference-relative agreement on \JointN{} of \SampleN{} files. Diagnostics
explain several causes of the remaining uncertainty, including import elision
and release-source association, without changing the \UnknownN{} UNKNOWN labels.
The supported outcome is an auditable enforcement contract and a bounded account
of its evidence, not a claim of complete architectural knowledge.

\section*{Artifacts, Author Responsibility and Disclosure}
The private author-review package provides the integrated source, selected
saved measurements, hashes, analysis and table-generation code, claim audit,
failure references and human-review routing. Earlier source snapshots, original
labels, failed attempts and complete phase packages remain preserved in the
research repository. The screened companion package supports saved-data checks;
it does not redistribute entire private or third-party repositories or claim
fresh-machine experimental replication. No public artifact URL, DOI or release
license is asserted.

OpenAI Codex assisted implementation, source inspection, experimental execution,
analysis and writing in the recorded phases. A separate non-authoring AI context
reviews the integrated manuscript and evidence package. This is internal
AI-assisted review, not independent human peer review or human annotation.
The author remains responsible for methods, citations, interpretations and
any eventual submission. There was no targeted external research funding,
and the author has declared no known competing interests. The affiliation does
not imply institutional endorsement.

\appendix
\section{Version History and Preserved Failures}\label{app:versions}
Paper revision, observer, reference and gate versions are separate identities.
O0 is the original lexical port; O1 is the isolated AST repair; O2 is the
supplementary tracked-package/cache-query observer; O3 includes the normalized
parent-main repair. G2 is the generic current-context layer and private epoch
store. The original reference remains immutable. The supplementary Babel/
TypeScript reference R2 was corrected to R3 after a binding-check audit; R4
repairs trailing-slash directory handling and R5 adds bounded exact-asset support.
The evidence manifest maps these records to the frozen input commit
\texttt{a140c2ae9a8a50ec8ad754bd947da2ead23c1d22}.

Under the original reference, 193/240 files qualified. The original port had
817 TP, one FP, zero FN and 192/193 exact files; the sensor ablation had
112 TP, zero FP, 705 FN and 55/193 exact files; native dependency-cruiser had
817 TP, seven FP, zero FN and 188/193 exact files. These are historical
reference-relative outcomes, not directly comparable accuracy trends.
At R3/O2, the joint subset contained 161 files and 687 TP. An earlier manuscript
incorrectly printed 787 TP; the eight saved strata sum to 687. The excess 100
was a writing-level arithmetic error, not a parser improvement. R5/O3 has
166 jointly exact files and 713 TP. Differing qualification populations prevent
treating 817, 687 and 713 as a single accuracy trajectory.

The original candidate full-tree unresolved counts were 189 for Fastify, one
for tRPC (plus a lexical issue), 131 for Vite and 54 for Vue. All 120 candidate
timing attempts failed qualification. The failure history also preserves the
nine false-PASS probes, acquisition failures, the first package pilot, both
114/120 qualification attempts, reference-binding omissions, shared-reporter
projection failure and same-process module-cache problem. Later successful
records are added with their own identities, not substituted for these attempts.

The diagnostic phases do not change R5/O3 qualification. V6's TypeScript 6.0.2
acquisition did not admit a compiler and remains a historical failure. V7
subsequently validated a complete archive before executing the recorded emission
witnesses. The acquisition phase's \texttt{compiler\_executed=false} does not
describe the later emission phase. There is no new compiler execution in V8.

\section{Diagnostic Details and Human Review}
The queue is routed by case identity, not by an inferred success label.
Forty JS/MJS cases retain a question about the intended source/type/runtime
relation. Five TypeScript type-test cases additionally require the effective
external compiler/TSTyche configuration. Three local-package cases require an
admissible installation context beyond copied-layout execution. Two generated
entry cases require a decision about source versus build artifacts; the legacy
case also retains a different attested source commit. One scaffold requires
generation options and installation context. The 23 UNKNOWN files outside the
queue have their own source-linked records and human questions.

For the distribution check, Fastify 5.12.7 contains 363 regular files, all
matching corresponding pinned-source bytes; the registry \texttt{gitHead}
matches its fixed commit. This is one-direction distribution-to-source
correspondence, not evidence that the entire repository was published. Across
Fastify, Vite and legacy, 409 regular archive members were checked, which is an
inventory total rather than 409 independent research cases. Vite has 39 regular
files and legacy seven; of 19 and three overlapping source files, respectively
18 and two match. Each differing overlap is \texttt{package.json}. The retained
manifest field differences are not normalized away.

Saved provenance verification uses Sigstore 4.1.0 with npm 11.9.0, exact
workflow identity and GitHub issuer expectations for SLSA provenance, official
registry-key verification for publish statements, and the recorded trust data.
Subject digests bind statements to the saved tarballs. Four targeted negative
controls tamper with the signed source payload or supply an incorrect expected
workflow identity; they test these checks, not
all possible forgeries. Vite's attested commit is
\texttt{284f6b58d6b511cebdd52ab7731975f385bf3c6b}; legacy's is
\texttt{48139b4828cbb8b6325aec229fa2a9be923d88a2}.
Neither build was reproduced locally. The recorded workflow invocation is a
statement field, not evidence that this study inspected every workflow log.

The R4/O3 directory and parent-main tests first exposed five failures; two
further exact-asset tests failed before R5. The final 41-test engineering suite
is regression evidence. The asset fallback applies only to explicit relative
CSS/JSON targets when TypeScript selects no target, with no extension guessing
and with containment, link and case-collision refusals retained. It establishes
a tracked literal relation, not runtime or bundler executability.

\section{Reproducibility and Evidence Map}\label{app:repro}
The private bundle's \texttt{evidence/manifest.json} records the scientific
input commit, Git blob identities and SHA256 hashes. Inputs are read from those
blobs, independently of the starting worktree-preservation baseline. The
normalizer rejects missing required fields and hash mismatches. Coverage selects
only the four repository-level rows; random and stress strata are checked
separately. The full UNKNOWN set is partitioned by case identities into the
51-case queue and 23 other files. TypeScript pairing uses source identity,
source hash and the verbatim option, checks unique one-to-one keys and compares
the saved output bytes.

\path{collect_evidence.py} reconstructs normalized metrics from saved phase
results. \path{render_tables.py} generates the numerical macros and nine
tables. \path{check_revision.py} checks source hashes, claims, generated
content, protected paths and package dependencies. These checks do not rerun
observers, compilers, native tools, functional tests or provenance verification.
The original experiment and phase-analysis code remains at the fixed private
repository commit; exact input selectors and retrieval instructions are included.
The screened reproduction bundle is an integration-reproduction package, not
a claim that every historical experiment is self-contained or publicly released.

The four external snapshots are Vue \texttt{bf5f0dd877dd},
Vite \texttt{284f6b58d6b5}, tRPC \texttt{fcc6bca0ece2}, and
Fastify \texttt{9f400035ebaf}; full hashes and sample rules remain in the locked
research design. Saved original E2 records preserve source anchors, constructed
families and N/A outcomes. V2 contains the binding, challenge, scoped-operation
and cost records; V3 supplies the current E1 reconciliation; V4 supplies complete
coverage and UNKNOWN routing; V5--V7 supply emission and artifact witnesses.
The separate claim ledger distinguishes observations, derived summaries,
assumptions, interpretations and limitations.

The LaTeX source archive contains only manuscript dependencies. It is compiled
after extraction into a separate directory without reading sibling research
files. The selected reproduction archive excludes operational caches, credentials,
whole repositories and unredacted machine logs. Original failures remain in the
research repository even when excluded from this screened archive. Author review
of relation semantics and redistribution rights remains necessary before any
public release.

\end{document}